\documentclass[12pt,osajnl2,preprint,showpacs]{revtex4}  
\usepackage{amsmath,amssymb,accents}

\begin{document}

\title{Ramsauer approach for light scattering on non-absorbing spherical particles and application to the Henyey-Greenstein phase function}


\author{Karim Louedec,$^{1,2,*}$ and Marcel Urban$^{2}$}
\address{$^1$Laboratoire de Physique Subatomique et de Cosmologie (LPSC), \\ UJF-INPG, CNRS/IN2P3, Grenoble, France}
\address{$^2$Laboratoire de l'Acc\'el\'erateur Lin\'eaire (LAL), \\ Univ Paris-Sud, CNRS/IN2P3, Orsay, France}
\address{$^*$Corresponding author: karim.louedec@lpsc.in2p3.fr}

\begin{abstract}
We present a new method to study light scattering on non-absorbing spherical particles. This method is based on the Ramsauer approach, a model known in atomic an nuclear physics. Its main advantage is its intuitive understanding of the underlying physics phenomena. We show that although the approximations are numerous, the Ramsauer analytical solutions describe fairly well the scattering phase function and the total cross section. Then this model is applied to the Henyey-Greenstein parameterisation of scattering phase function to give a relation between its asymmetry parameter and the mean particle size.   
\end{abstract}


\maketitle 

The subject of light scattering by small particles is present in several scientific areas such as astronomy, meteorology and biology~\cite{Hulst,Bohren}. The first model by Lord Rayleigh in 1871 dealt with light scattering by particles whose dimensions are small compared to the wavelength. A significant improvement came with the Mie solution~\cite{Mie} which describes light scattering by spherical particles of any size. In an apparently different domain, the scattering of a low energy electrons by atoms studied by Ramsauer experimentally~\cite{Ramsauer} showed surprising structures, and it was several years before the explanation found by Bohr describing the electron as a plane wave~\cite{e-atome1,e-atome2,Karwasz}! The Ramsauer effect was certainly the first phenomenon showing the wave properties of matter. This framework has since then been used to describe very different types of collisions such as atom-atom~\cite{atome-atome1} or even neutron-nucleus~\cite{neutron-noyau1,neutron-noyau2,neutron-noyau3,Peterson,Abfalterer}. Our idea is that since light behaves like a wave, it should be also possible to apply Bohr/Ramsauer's ideas to light scattering by spherical particles.

Sect.~\ref{sec:light_scat} is a brief introduction of some quantities concerning light scattering on (non-absorbing) spherical particles. Sect.~\ref{sec:ramsauer} describes the Ramsauer effect and provides details on the way to obtain the Ramsauer scattering phase function and its simplified solution. We compare the Ramsauer predictions with the Mie calculations, emphasizing certain characteristics of light scattering. Then, in Sect.~\ref{sec:totalCS}, we integrate the differential cross section to obtain the total cross section and the associated extinction efficiency parameter using the Ramsauer approach. Finally, Sect.~\ref{sec:HenyeyGreenstein} presents an application of the Ramsauer approach for the Henyey-Greenstein parameterisation of the scattering phase function; a basic relation between its asymmetry parameter and the mean particle size is obtained.

\section{The Mie predictions for scattering of light over a sphere of non-absorbing dielectric}
\label{sec:light_scat}
When light interacts with particles, two different kinds of processes can occur. The energy received can be reradiated by the particle at the same wavelength. Usually, the reradiation takes place with different intensities in different directions. This process is called {\it scattering}. Alternatively, the radiant energy can be transformed into other wavelengths or other forms of energy, such as heat: {\it absorption} takes place. In the following we will consider absorption equal to zero. The intensity of light scattered at an angle $\theta$ and a distance $r$ by a single spherical particle with a radius $R$, from a beam of unpolarised light of wavelength $\lambda$ and intensity $I_0$ is given by
\begin{equation}
I(\theta) = \frac{I_0}{r^2}\times \sigma_{\rm sca}(R,\lambda)\times P(\theta),
\label{eq:01}
\end{equation}
where $I_0$ is the initial intensity of the beam, $\sigma_{\rm sca}(R,\lambda)$ the scattering cross section and $P(\theta)$ the scattering phase function. The integral of $P(\theta)$ over all solid angles has to be equal to unity. The scattering cross section $\sigma_{\rm sca}(R,\lambda)$ has the dimension of area but is not in general equal to the particle cross-sectional area. Indeed it is customary to define a scattering efficiency
\begin{equation}
Q_{\rm sca} = \frac{\sigma_{\rm sca}(R,\lambda)}{\pi R^2},
\label{eq:02}
\end{equation}
where $\pi R^2$ is the geometric cross section in the case of a sphere of radius $R$. The scattering efficiency represents the ratio of the energy scattered by the particle to the total energy in the incident beam intercepted by the geometric cross section of the particle. Since absorption is assumed null here, the extinction efficiency $Q_{\rm ext}$ is equal to $Q_{\rm sca}$ and the corresponding total cross section $\sigma_{\rm tot}$ is equal to $\sigma_{\rm sca}$.

As already mentioned, the ratio of the particle size to the wavelength of the radiation of interest is of
crucial importance for the particle's optical properties as well as for the choice of a suitable method for calculating
those properties. We therefore define the dimensionless size parameter as
\begin{equation}
x = \frac{2\pi R}{\lambda} = k_{\rm out}\,R,
\label{eq:03b}
\end{equation}
where $R$ is the radius of a spherical particle, $\lambda$ the light wavelength and $k_{\rm out}$ the wave number outside the sphere. In the case of non-spherical particles, $R$ might represent the radius of a sphere having the same volume. The theory describing the scattering of an electromagnetic plane wave by a homogeneous sphere of arbitrary size was originally presented by Gustav Mie~\cite{Mie}. Particles with a size comparable to the wavelength of visible light are relatively common in nature. In this case, the Rayleigh theory is not applicable anymore because the field is not uniform over the entire particle volume. The laws describing the total scattered intensity as a function of the incident wavelength and the characteristics of the particle are much more complex than those for Rayleigh scattering. 

\subsection{Extinction efficiency: the total cross section}
\label{sec:ext_eff}
The $1/\lambda ^4$ dependence of the total scattered intensity found by Rayleigh is no longer valid generally for the Mie solution. The extinction efficiency factor  $Q_{\rm ext} = \sigma_{\rm tot}/(\pi R^2) $ depends on the radius $R$ and on the relative refractive index of the particle $n$. The relative refractive index is defined as the ratio of the refractive index of the particle over the refractive index of the medium. In the following, where the medium is air, the relative refractive index corresponds to the refractive index of the particle $n$. The Mie theory uses Maxwell's equations to obtain a wave propagation equation for the electromagnetic radiation in a three dimensional space, with appropriate boundary conditions at the surface of the spherical particle. The extinction efficiency factor obtained is
\begin{equation}
\label{eq:A1}
Q_{\rm ext} = \frac{2}{x^2} \sum_{\ell=1}^{\infty} (2 \ell+1) \mathop{\mathrm{Re}} (a_{\ell}+b_{\ell}),
\end{equation}
where the Mie scattering coefficients $a_{\ell}$ and $b_{\ell}$ are functions of the size parameter $x = 2 \pi R/ \lambda = k_{\rm out}R$ and of the index of refraction $n$~\cite{Hulst,Bohren}. To compute $Q_{\rm ext}$ numerically using Eq.~(\ref{eq:A1}), it is necessary to truncate the series, keeping enough terms to obtain a sufficiently accurate approximation. The criterion developed by Bohren in~\cite{Bohren} was obtained through extensive computation. Analysis of the convergence behaviour of Eq.~(\ref{eq:A1}) revealed that the number of required terms $N$ was slightly influenced by refractive index and had to be the closest integer to $x + 4x^{1/3} + 2$~\cite{Wiscombe}. For instance, for a raindrop of $50\, \mu$m radius and a visible wavelength of $0.6\, \mu$m, the number of required terms is $N=558$. Nowadays, computers have reached a point where computing time is no longer a major problem. The closed form Ramsauer approach, which we will develop later, would have been useful some years ago to provide computational ease. In the present context it furnishes useful insights into the different physical processes involved.

\begin{figure}[p]
\centering
	\includegraphics [width=4.25in] {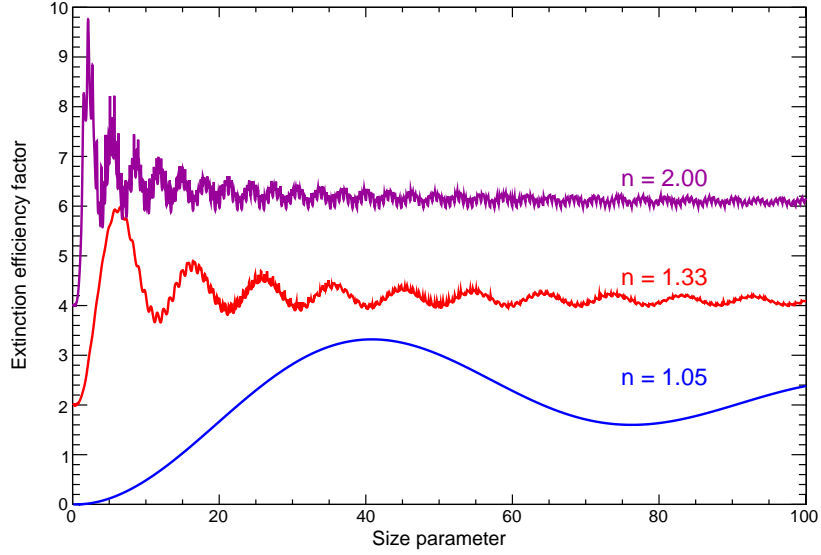}
\caption{{(Color online) The extinction efficiency factor $Q_{\rm ext}$ versus the size parameter $x$, for non-absorbing spherical particles with relative refractive indices $n=1.05,\: 1.33$, and $2.00$}. $x$ is given by the relation $x = 2 \pi R/ \lambda = k_{\rm out}R$. The vertical scale applies only to the lowest curve, the others being successively shifted upward by 2.}
\label{fig:Mie_QextfSizeParameter}
\end{figure}

\begin{figure}[p]
\centering
	\includegraphics [width=4.25in] {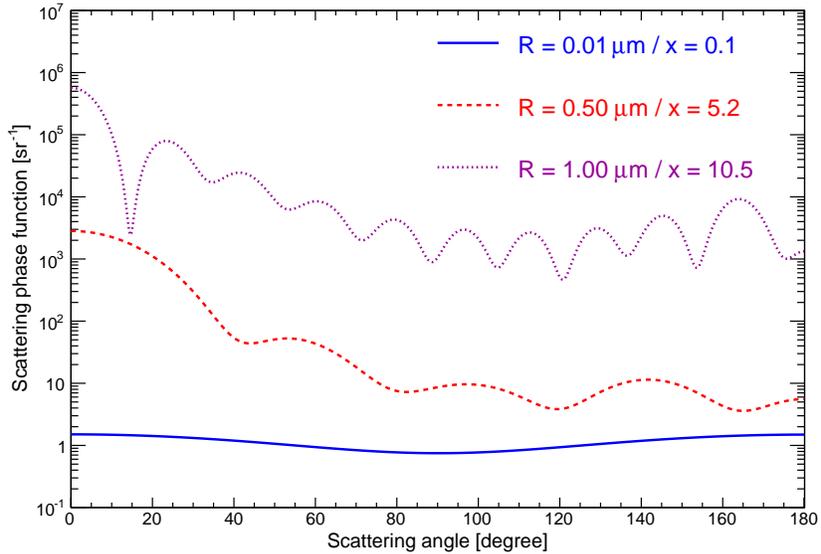}
\caption {{(Color online) Angular scattering for water droplets illuminated by unpolarised light, for three different radii.} The incident wavelength is fixed at $0.6~\mu$m. The vertical scale applies only to the lowest curve, the others being successively multiplied by 100.}
\label{fig:Mie_PhaseFunctions}
\end{figure}

Fig.~\ref{fig:Mie_QextfSizeParameter} shows, for three values of the index of refraction, $Q_{\rm ext}$ as a function of the size parameter $x$. Each curve is characterised by a succession of maxima and minima with superimposed ripples. The amplitude of the large oscillations and of the ripples grows with $n$. For particles much larger than the wavelength of the incoming light, i.e.\ with a size parameter $x \gg 1$, it might be expected from geometrical optics that the total cross section would be $\pi R^2$, and thus $Q_{\rm ext}$ would approach $1$. However, Fig.~\ref{fig:Mie_QextfSizeParameter} shows an interesting and somewhat puzzling trend, the scattering efficiency approaching $2$. It is the so-called {\it extinction paradox}, and is discussed in detail by Van de Hust~\cite{Hulst}. The phenomenon is apparent only for observations made far from an object, so that light that is scattered even at a small angle can be considered removed from the beam. The total cross section results from the geometrical contribution (classical limit) of $\pi R^2$ and an equal contribution from the diffraction of the incident plane wave at the surface of the sphere. The diffraction contribution is strongly peaked forward. For nearby macroscopic objects, this diffracted light is not distinguishable from unscattered light at $\theta = 0^\circ$ and the paradox is not observed.

\subsection{Angular scattering: the differential cross section}
\label{sec:ang_scat}
Mie scattering by single particles irradiated by laser sources is sufficiently strong to be detected with high signal-to-noise ratios for particles larger than about $0.1~\mu$m. The noise results from Rayleigh scattering by residual molecules and from the electronics used. The recorded signal depends on the scattering angle, as well as on the particle size and refractive index~\cite{ExpMie}. The angular dependence can be calculated from Mie theory. In this study, we use a code developed by Philip Laven~\cite{Laven}. For values of the size parameter $x$ approaching unity, an asymmetry favouring forward scattering appears. For $x\gg 1$, forward scattering increases even more strongly, showing very rapid changes for small increases in the scattering angle $\theta$. Some of this behaviour is shown in Fig.~\ref{fig:Mie_PhaseFunctions} for water droplets of different sizes when illuminated by unpolarised light of $\lambda = 0.6~\mu$m. Very small droplets show typical Rayleigh behaviour with a symmetric curve and a weak minimum at $90^\circ$. When the size parameter $x$ increases, additional minima and maxima appear and a strong asymmetry develops with the forward scattering several orders of magnitude stronger than the backscattering.

\begin{figure}[t]
\centering
	\includegraphics [width=4.25in]{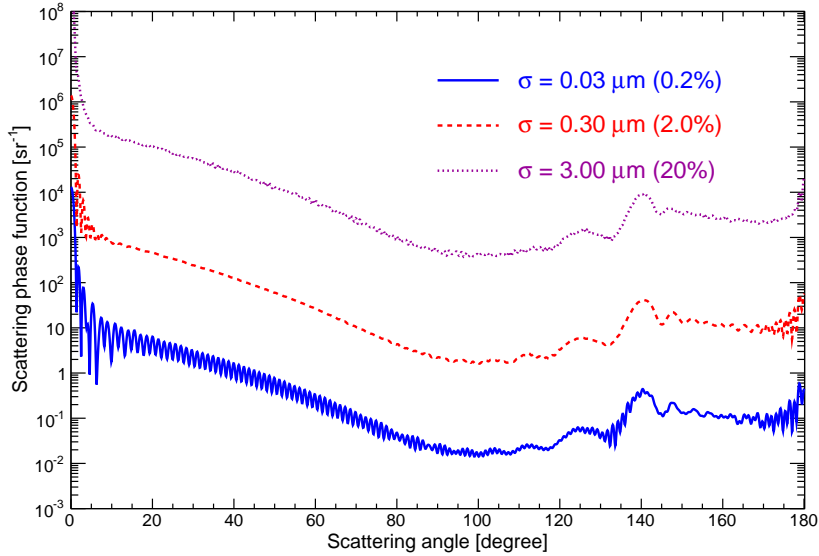}
\caption {{(Color online) Angular scattering for a distribution of water droplets illuminated by unpolarised light, for three different standard deviations.} The mean radius $\bar{R}$ is fixed at $15~\mu$m and the incident wavelength at $0.6~\mu$m. The vertical scale applies only to the lowest curve, the others being successively multiplied by 100.}
\label{fig:Mie_LogNPhaseFunctions}
\end{figure}

Most natural and artificial collections of spherical particles do not exhibit the large fluctuations as in Fig.~\ref{fig:Mie_PhaseFunctions}. Even a narrow dispersion in the distribution of the particle size washes out these features that strongly depend on size. The size distribution can be described using a log-normal function defined as follows
\begin{equation}
n(R |\bar{R},\sigma) = \frac{{\rm d} N(R)}{{\rm d} R} = \frac{N}{\sqrt{2\pi}\;\log_{\rm 10}\, \sigma} \frac{1}{R} \exp\left(-\frac{\log^2_{\rm 10}(R/\bar{R})}{2\; \log^2_{\rm 10}\sigma} \right),
\label{SizeDistr:eq0}
\end{equation}
where $R$ is the radius of a spherical particle, $\sigma$ is the geometric standard deviation, $\bar{R}$ is the geometric mean radius, and $N$ is the total number of particles. Phase functions for three different geometric standard deviations are plotted in Fig.~\ref{fig:Mie_LogNPhaseFunctions}, with a mean radius $\bar{R}$ fixed at $15~\mu$m and a wavelength equal to $0.6~\mu$m. As expected, a smaller geometric standard deviation implies a wavier scattering phase function. The ripples are stronger in the forward and backward scattering peaks. Consequently, in nature, where particle radii are always dispersed, light scattering will be described using smoothed phase functions (see Sect.~\ref{sec:HenyeyGreenstein}).


\section{Application of the Ramsauer effect for light scattering}
\label{sec:ramsauer}
\begin{figure}[t]
\centering
 \includegraphics[width=6.5in]{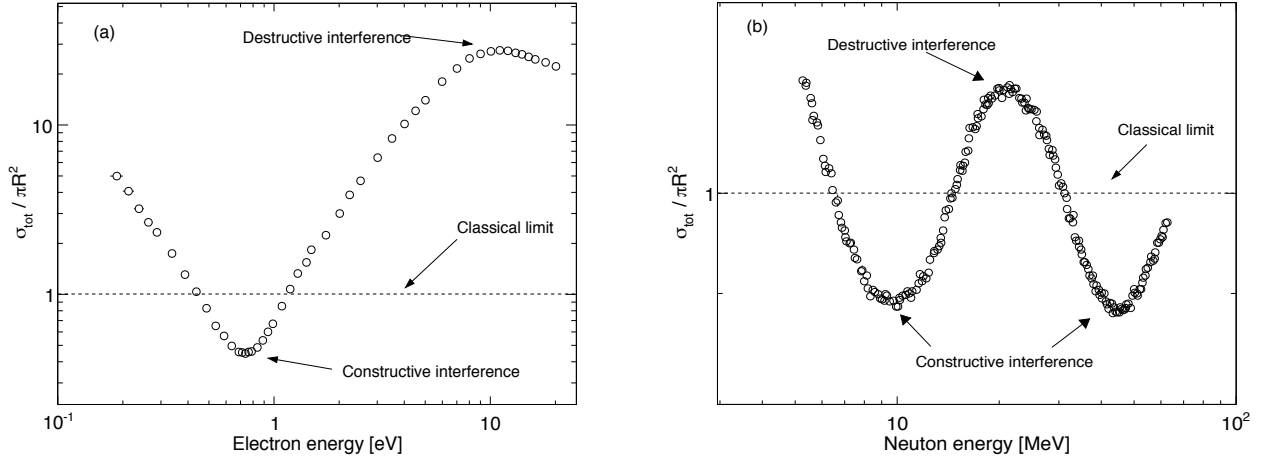}
\caption{{(a)~Electron-krypton normalised total cross section versus energy \cite{Karwasz}. (b)~Normalised neutron-lead total scattering cross section versus energy \cite{Abfalterer}.}}
\label{fig:Ramsauer_data}
\end{figure}

The Ramsauer effect was discovered in 1921~\cite{Ramsauer} while studying electron scattering on argon atoms. The total cross section versus the electron energy showed a surprising dip around $1$\,eV. In Fig.~\ref{fig:Ramsauer_data} we see recent measurements of electron scattering over krypton and neutron scattering on lead nuclei.

\subsection{The Ramsauer scattering phase function}
\begin{figure}[t]
\centering
\includegraphics{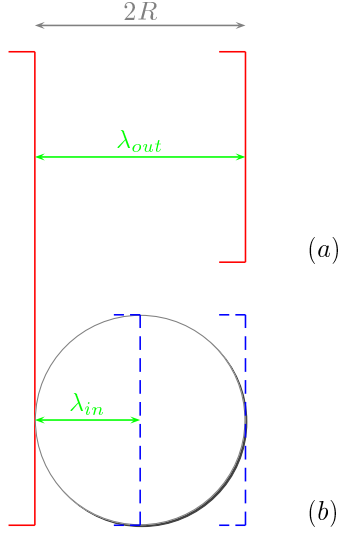}
\caption{{(Color online) Qualitative picture of the Ramsauer phenomenon.} The wavelength of the light is supposed to be reduced in the dielectric. This picture shows the case where the contraction of the wavelength between (a) outside and (b) inside the medium (dashed line) is such that they come out in phase. Thus the sphere looks invisible resulting in an almost zero cross section.}
\label{fig:2}
\end{figure}

\begin{figure}[t]
\centering
\includegraphics[width=3.4in]{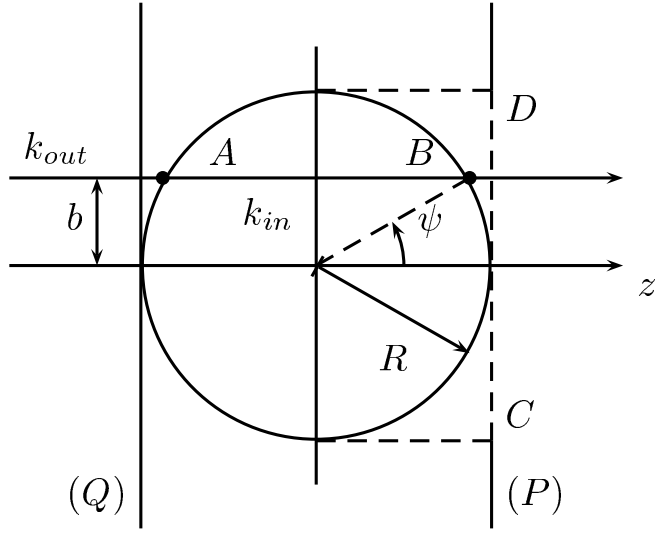}
\caption{{Definition of the variables used in the text to calculate the phase function for light scattering by a spherical particle of radius $R$.} $k_{\rm in}$ and $k_{\rm out}$ are the wavenumbers for the light inside and outside, respectively, and $b$ is the impact parameter.}
\label{fig:3}
\end{figure}

The idea in modelling the phenomenon is to consider the incident particle as a plane wave with one part going through the target and another which does not (Fig.~\ref{fig:2}). The two parts recombine behind the target and then interfere with each other, producing the oscillating behaviour. Depending upon the impact parameter $b$ (Fig.~\ref{fig:3}), light rays going through the drop accumulate a phase shift.

Using the Huygens-Fresnel principle, the propagation of light can be described with the help of virtual secondary sources. Each point of a chosen wavefront becomes a secondary source and the light amplitude at any chosen location is the integral of all these sources emitting spherical waves towards that observation point. The secondary sources are driven by the incident wave. Their amplitude is
\begin{equation}
-\frac{i}{\lambda} \frac{(1+\cos\theta)}{2} A_{\rm incident} {\rm d} S
\end{equation}
where $\lambda$ is the wavelength, $A_{\rm incident}$ is the amplitude of the incoming light, $\theta$ is the angle
between the direction of the incident light and the direction from the virtual source to the observation point, and ${\rm d} S$ is the elementary area of the secondary source stands.

We define the plane P, shown in Fig.~\ref{fig:3}, as the location of our secondary sources. The shadow of the sphere on the plane P is the disk CD. If the plane Q is chosen as the origin of the phases, the incident light is a plane wave which, on the plane P, has the amplitude $e^{i 2 k_{\rm out} R}$. The amplitude distributions of the secondary sources can be split into two parts: an undisturbed plane wave and a perturbation in the CD region only. The undisturbed initial plane wave, when integrated, is described by a Dirac delta function in the forward direction. Thus the differential cross section comes from the sources on CD. The observation points are considered to be very far away so, to get the amplitude at an angle $\theta$, we just sum all directions parallel to $\theta$.

Let $\rho$ and $\phi$ be the polar coordinates in the plane P. The amplitudes from the virtual sources in the disk CD are then
\begin{equation}
\begin{split}
A_{\rm disk}(\rho)&= -\frac{i}{\lambda} \left[-e^{i 2k_{\rm out} R} + e^{i n 2 k_{\rm out} \sqrt{R^2-\rho^2}}\, e^{i k_{\rm out} (2R-2\sqrt{R^2-\rho^2})} \right] \frac{1+\cos\theta}{2}\\
&=i \frac{k_{\rm out}}{2 \pi}e^{i 2 k_{\rm out} R}\left[1- e^{i 2 k_{\rm out} \sqrt{R^2-\rho^2}\;(n-1)}\right] \frac{1+\cos\theta}{2} .
\end{split}
\label{eq:C1}
\end{equation}

The resulting scattering amplitude $f_{\rm sphere}(\theta)$ is obtained after moving the origin of the phases to the plane P. This is simply done through a multiplication by $e^{-i 2 k_{\rm out} R}$
\begin{equation}
\begin{split}
f_{\rm sphere}(\theta)&=e^{-i 2 k_{\rm out} R} \int_0^{2\pi} {\rm d} \phi \int_0^R A_{\rm disk}(\rho) \, \rho \, e^{i k_{\rm out} \rho \cos\phi\sin\theta} {\rm d} \rho \\
&= i \frac{k_{\rm out}}{2 \pi}\frac{1+\cos\theta}{2} \int_0^{2\pi} {\rm d} \phi \int_0^R e^{i k_{\rm out} \rho \cos\phi\sin\theta} \left[1- e^{i 2 k_{\rm out} \sqrt{R^2-\rho^2}\;(n-1)}\right]\,\rho \, {\rm d} \rho ,
\end{split}
\label{eq:C2}
\end{equation}
where the term $\rho \cos\phi\sin\theta$ represents the path length difference between a source in the disk and a source at the disk centre for an observer placed at an infinite distance from the disk.

The integral over the angle $\phi$ is a Bessel function
\begin{equation}
J_0(u) = \frac{1}{2 \pi} \int\limits_0^{2\pi} e^{i u \cos \phi} {\rm d} \phi .
\label{eq:C3}
\end{equation}

Thus Eq.~(\ref{eq:C2}) becomes
\begin{equation}
f_{\rm sphere}(\theta)= i k_{\rm out}\frac{1+\cos\theta}{2} \int\limits_0^R \rho J_0(k_{\rm out} \rho \sin \theta) \left[1- e^{i 2 k_{\rm out} \sqrt{R^2-\rho^2}\;(n-1)}   \right] {\rm d} \rho .
\label{eq:C4}
\end{equation}
This integral has no analytical solution. The Ramsauer differential cross section is 
\begin{equation}
\left[\frac{{\rm d} \sigma}{{\rm d} \theta}\right]_{\rm sphere}=\left| f_{\rm sphere}(\theta)\right|^2 ,
\label{eq:C5}
\end{equation}
and we deduce directly the Ramsauer phase function 
\begin{equation}
P_{\rm R}(\theta) = \frac{1}{\sigma}\left[\frac{{\rm d} \sigma}{{\rm d} \theta}\right]_{\rm sphere}=\frac{\left| f_{\rm sphere}(\theta)\right|^2}{\int\limits_0^{2\pi}{\rm d} \Phi \int\limits_0^\pi  {\rm d} \theta \sin \theta \left| f_{\rm sphere}(\theta)\right|^2}.
\label{eq:C5bis}
\end{equation}
This solution has to be computed numerically. Thus, this illustrates that predicting the behaviour of light scattering is not easy.

\subsection{The simplified Ramsauer scattering phase function}
\label{sec:Ramsauer_phasefunction}
We can get a formula in closed form if the sphere is approximated by a cylinder of radius $R$ and of height $2R$ along $z$. Eq.~(\ref{eq:C4}) simplifies into
\begin{equation}
\begin{split}
f_{\rm cyl}(\theta)&=i k_{\rm out} \frac{1+\cos\theta}{2} \int_0^R \rho J_0(k_{\rm out} \rho \sin \theta) \left[1- e^{i 2 k_{\rm out} R\,(n-1)}\right] {\rm d} \rho  \\
&= i k_{\rm out} \frac{1+\cos\theta}{2} \left[1- e^{i 2 (n-1)k_{\rm out} R}\right]  \int_0^R \rho J_0(k_{\rm out} \rho \sin \theta) {\rm d} \rho \\
&= i \frac{1+\cos\theta}{2} e^{i \frac{2 (n-1)\,k_{\rm out} R}{2}} (-2 i)\sin \left[\frac{2 (n-1)\,k_{\rm out}R}{2} \right] \frac{R}{\sin \theta} J_1(k_{\rm out} R \sin \theta),
\end{split}
\label{eq:C6}
\end{equation}
since the integral can be simplified using the fact that 
\begin{equation}
\int\limits_0^R u J_0(u) {\rm d} u = R \, J_1(R).
\end{equation}

The differential cross section is
\begin{equation}
\left[\frac{{\rm d} \sigma}{{\rm d} \theta}\right]_{\rm cyl}=\left| f_{\rm cyl}(\theta)\right|^2 = R^2\left[k_{\rm out}R \frac{1+\cos \theta}{2} \sin\left[(n-1)k_{\rm out}R \right]\frac{2 J_1(k_{\rm out}R\sin \theta)}{k_{\rm out}R \sin \theta} \right]^2.
\label{eq:C7}
\end{equation}
In fact, in the Ramsauer approach, the total cross section for a cylinder of radius $R$ and of height $L$ is given by
\begin{equation}
\sigma_{\rm tot}^{\rm cyl} = \int\limits_0^{2\pi}{\rm d} \Phi \int\limits_0^\pi  {\rm d} \theta \sin \theta \left[\frac{{\rm d} \sigma}{{\rm d} \theta}\right]_{\rm cyl} = 4 \pi R^2 \sin^2\left[\frac{(n-1)}{2}k_{\rm out}L \right] = 4 \pi R^2 \sin^2\left[(n-1)k_{\rm out}R \right],
\label{eq:C8}
\end{equation}
where $L=2R$ in our case. Thus, the differential cross section can be expressed as 
\begin{equation}
\left[\frac{{\rm d} \sigma}{{\rm d} \theta}\right]_{\rm cyl} = \frac{1}{4 \pi}\;\sigma_{\rm tot}^{\rm cyl} \left[k_{\rm out}R \frac{(1+\cos \theta)}{2} \frac{2 J_1(k_{\rm out}R\sin \theta)}{k_{\rm out}R \sin \theta} \right]^2 .
\label{eq:C9}
\end{equation}

As we will see in Sect.~\ref{sec:validity}, it is possible to use a similar equation for the differential cross section with a sphere, i.e.\
\begin{equation}
\left[\frac{{\rm d} \sigma}{{\rm d} \theta}\right]_{\rm sphere} = \frac{1}{4 \pi}\;\sigma_{\rm tot}^{\rm sphere} \left[k_{\rm out}R \frac{(1+\cos \theta)}{2} \frac{2 J_1(k_{\rm out}R\sin \theta)}{k_{\rm out}R \sin \theta} \right]^2 .
\label{eq:C11}
\end{equation}

In this subsection, we have shown that in the simplified Ramsauer approach (SR), the total cross sections for a cylinder and for a sphere can be different but the scattering phase functions coincide and are given by the relation
\begin{equation}
P_{\rm SR}(\theta) = \frac{1}{\sigma} \frac{{\rm d} \sigma}{{\rm d} \theta} = \frac{1}{4 \pi}\;\left[k_{\rm out}R \frac{(1+\cos \theta)}{2} \frac{2 J_1(k_{\rm out}R\sin \theta)}{k_{\rm out}R \sin \theta} \right]^2 .
\label{eq:C12}
\end{equation}
It seems interesting to note that this simplified phase function is independent of the relative refractive index. The validity of this independence will be verified in Sect.~\ref{sec:validity}. It has been assumed in this last part a cylinder of radius $R$ and of height $L=2R$, i.e. a volume equal to $2\pi R^3$. But, in the case of a sphere, the volume of the scattering center is different and equal to $(4/3)\,\pi R^3$. However, since the differential cross section is normalised to the total cross section to get the phase function, this difference is avoided.

As can be seen in Eq.~(\ref{eq:C12}), the Ramsauer approach is much more intuitive than Mie model. We note two properties of the Ramsauer phase function:
\begin{enumerate}
\item The amount of light scattered in the forward direction ($\theta = 0^\circ$) is proportional to the target area, i.e.\ square of the target radius: at constant wavelength, a larger sphere scatters more light in the forward direction, in agreement with the Mie prediction.
$$
\frac{1}{\sigma}\frac{{\rm d}\sigma}{{\rm d}\theta}(\theta=0^\circ) \propto R^2.
$$
\item The full width at half maximum (FWHM) of the forward scattering peak is proportional to the inverse of $k_{\rm out}R$.
$$
\Delta\theta_{\rm FWHM} = \frac{2}{k_{\rm out}R}.
$$
\end{enumerate}

\begin{figure}[p]
\begin{center}
 \includegraphics[width=4.25in]{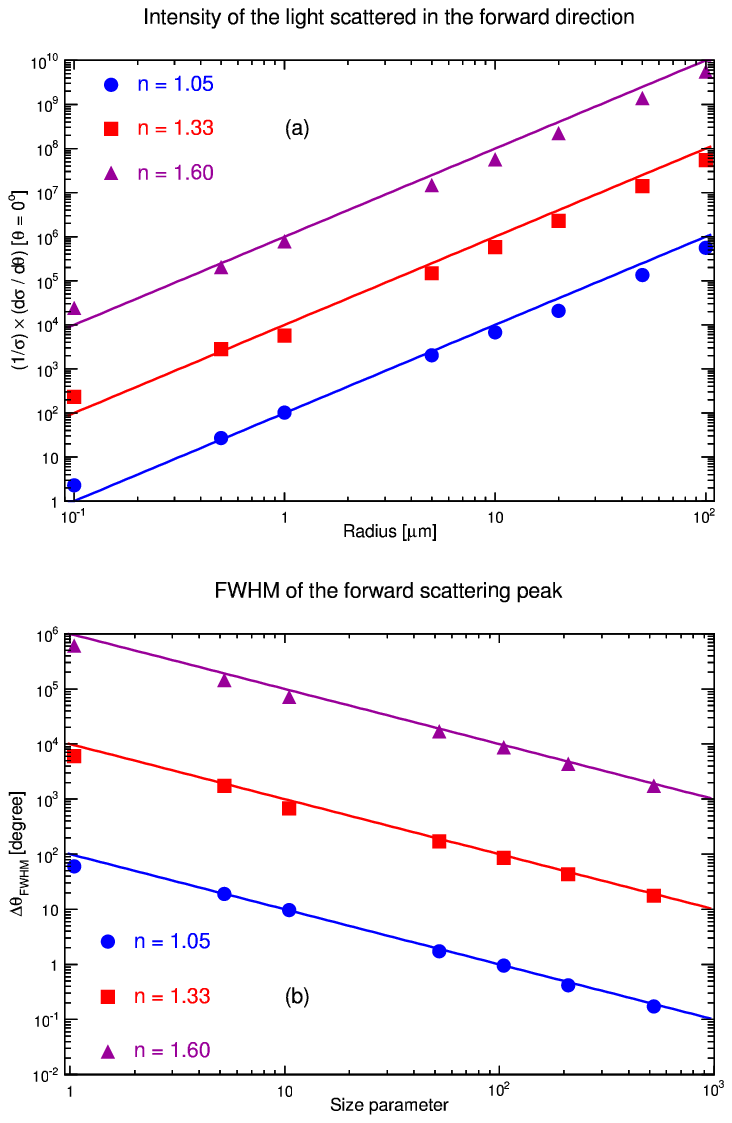}
\caption{{(Color online) Characteristic behaviours of physics quantities from the phase function.} The curves are obtained by Mie calculations (using~\cite{Laven}) for an incident wavelength of $0.6~\mu$m. Three different refractive indices are shown: $n=1.05$ (blue circles), $n=1.33$ (red squares) and $n=1.60$ (magenta triangles). The vertical scale applies only to the lowest curve, the others being successively multiplied by $100$.}
\label{fig:MieIntuitivePhaseFunction}
\end{center}
\end{figure}

\begin{figure}[p]
\begin{center}
\includegraphics[width=4.25in]{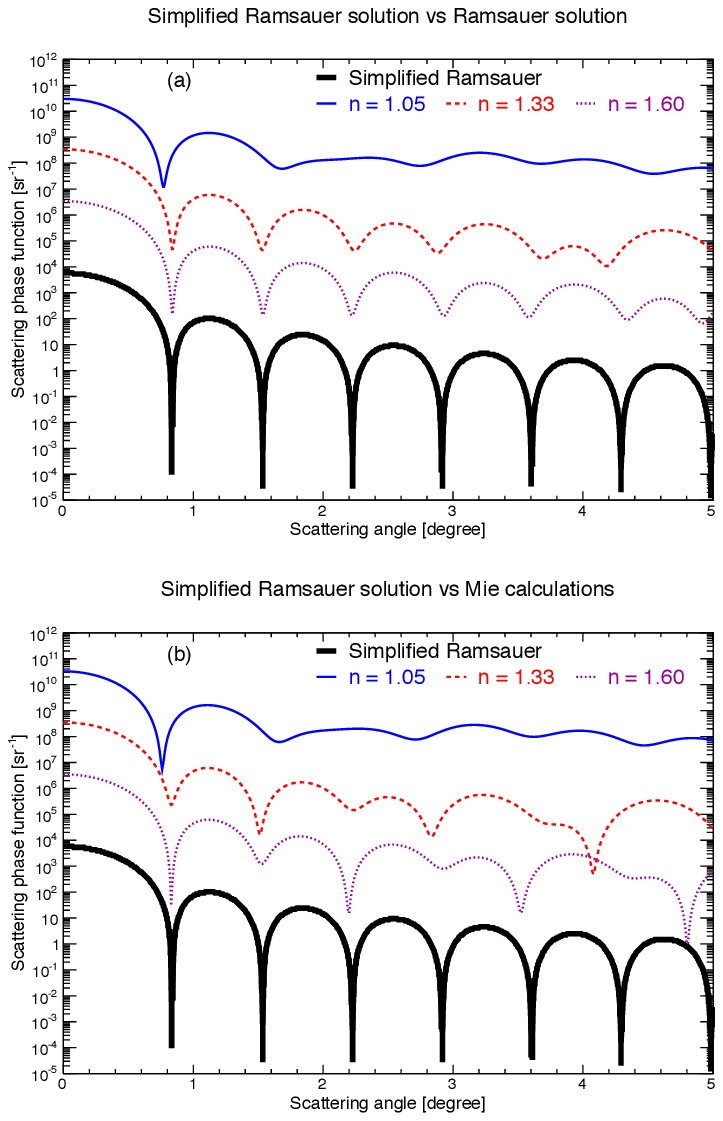}
\caption{{(Color online) The simplified Ramsauer solution, independent of the refractive index, compared to the Ramsauer solution and Mie calculations.} (a)~SR vs Ramsauer. (b)~SR vs Mie theory. Three different refractive indices are shown: $n=1.05$, $1.33$ and $1.60$. The simplified solution, independent of the refractive index, is in black thick line. The incident wavelength is fixed at $0.6~\mu$m and the sphere radius is equal to $10~\mu$m. The vertical scale applies only to the lowest curve, the others being successively multiplied by $100$.}
\label{fig:RamsauerVsMie}
\end{center}
\end{figure}

By detailed Mie calculations, these phase function properties deduced from the analytical Ramsauer solution can be verified. Fig.~\ref{fig:MieIntuitivePhaseFunction} gives the phase function for pure forward scattering as a function of $R$ and the full width at half maximum as a function of the size parameter. For three different refractive indices, the observed behaviour is confirmed by the Mie calculations. However, an analysis comparing the mean radii estimated by the Ramsauer approach with the true values used in Mie calculations has stressed an overestimation of around $30\%$ of the values. This overestimation is independent of the incident wavelength and of the relative refractive index. After all, whereas it is difficult to obtain clear physical interpretations with Mie theory, the solutions being in the form of infinite series, the simplified Ramsauer solution makes easier prediction of light scattering behaviour possible.

\subsection{Validity of the Ramsauer solutions}
\label{sec:validity}
From the Ramsauer approach, two scattering phase functions were obtained: one from an exact calculation, $P_{\rm R}(\theta)$, and the second resulting from an approximation but given by a closed formula, $P_{\rm SR}(\theta)$. The goal of this section is to evaluate the two solutions with respect to the Mie theory, taken here as the reference. As noted above, the simplified solution does not depend on the refractive index of the target. It is thus important to see how the Mie calculations and the Ramsauer solution evolve with respect to the refractive index. Fig.~\ref{fig:RamsauerVsMie} shows the phase functions obtained from (a) the Ramsauer solution and (b) Mie theory for different refractive indices. For comparison, the one obtained from the simplified Ramsauer formula is also plotted with a thick black line. Although the phase function is almost independent of the refractive index for high $n$ values, this is not the case when $|n -1|$ approaches $0$. Moreover, the Ramsauer solution is in better agreement with the Mie curves when the refractive index is low.

Consequently, when we want to study light scattering phenomena for spherical particles, as a first approximation, we can develop models without knowing the refractive index of the particles. This is due to the fact that Fraunhofer diffraction is mostly responsible for small-angle scattering and it is nearly independent of the relative refractive index. Thus, forward scattering is sometimes favoured in the design of optical particle counters to eliminate the effect of refractive index on the measurement of particle size.

The different expectations deduced from the Ramsauer approach could be also checked by a basic experimental setup. One of the instruments for angular light scattering measurements is the polar nephelometer~\cite{Barkey,Castagner}. It consists of a collimated light source (usually a laser) and an arm that can be rotated around the sample of scattering particles. Two kinds of light collectors could be used: a single detector adjustable to different angles, or a set of multiple fixed detectors positioned around the sample and collecting simultaneously the light scattered at different angles. If the main goal is to get the best angular resolution for the scattering phase function, the first solution would be employed. The sample could be composed of polystyrene beads in suspension in water. Different mean sizes and different size dispersions are available in~\cite{Aaldrich}. We note that a detector linearity is required over a large intensity range. Indeed, as shown in Fig.~\ref{fig:Mie_PhaseFunctions}, a high intensity difference can be probed between the forward scattering peak and the rest of the angular range. Using the experimental setup briefly presented here, it seems possible to check the different theoretical expectations deduced by the Ramsauer approach. However, this work is beyond the scope of this study but should be done in future.

\section{The Ramsauer total cross section and extinction efficiency}
\label{sec:totalCS}
Since the results of the phase function predicted by the Ramsauer approach seem in agreement with the Mie calculations, this section will check if this is also the case for the total cross section. Two different methods will be presented to get the cross section by the Ramsauer approach: the first one based on the optical theorem (and independent of the results given in Sect.~\ref{sec:ramsauer}), and the second by integrating the differential cross section over ($\theta$, $\phi$).

\subsection{Ramsauer total cross section via the Optical theorem}
In scattering theory, the wave function far away from the scattering region must have up to the normalisation factor the form of a plane wave and an out-going spherical wave,
\begin{equation}
\label{eq:B1}
\Psi (\vec{r}) =  e ^{i \vec{k}_{\rm out} \vec{r}} + f(\theta) \frac{e ^{i k_{\rm out} r}}{r}.
\end{equation}

The optical theorem asserts that
\begin{equation}
\label{eq:B2}
\sigma _{\rm tot}=\frac{4\pi}{k_{\rm out}}\mathop{\mathrm{Im}}f(\theta = 0),
\end{equation}
where $f(\theta = 0)$ is the forward scattering amplitude. Under the approximation for the scattering of a scalar (spinless) wave on a spherical and symmetric potential, the scattering amplitude at a given polar angle $\theta$ can be written as a sum over partial wave amplitudes, each of different angular momentum $\ell$,
\begin{equation}
\label{eq:B3}
f(\theta)=\frac{1}{2 i k_{\rm out}}\sum_{\ell =0}^\infty(2 \ell +1)P_{\ell} (\cos\theta)\left[\eta_{\ell}\, e ^{2 i \delta_{\ell}}-1\right],
\end{equation}
where $\eta_{\ell}$ is the inelasticity factor ($\eta_{\ell} = 1$ in our case of a non-absorbing sphere), $\delta_{\ell}$ is the phase shift ($\delta_l$ is real for pure elastic scattering), and $P_{\ell}(\cos\theta)$ is the Legendre polynomial.

From Fig.~\ref{fig:3}, $\delta_{\ell}=(k_{\rm in}-k_{\rm out})R\cos\psi$. In the approximation of forward scattering, $P_\ell(\cos\theta)\rightarrow 1$. From Bohr momentum quantisation applied to the impact parameter $b=R \sin\psi$ (see Fig.~\ref{fig:3}), we have $b\,p=\ell \, \hbar$ and $p=\hbar \,k_{\rm out}$ with $\ell=b \,k_{\rm out}$. By substituting the discrete sum over $\ell$ by an integral over $\ell$ or over the impact parameter 
\begin{equation}
\sum_{\ell} \rightarrow \int\limits {\rm d} \ell \rightarrow k_{\rm out}\int\limits {\rm d} b ,
\end{equation}
the forward scattering amplitude can be written in terms of the impact parameter

\begin{equation}
\label{eq:B4}
f(\theta = 0)=\frac{k_{\rm out}}{i} \int\limits_{0}^R \left( e ^{i 2(k_{\rm in}-k_{\rm out})R\cos \psi}-1\right) b \, {\rm d} b ,
\end{equation} 
or in terms of the angle $\psi$
\begin{equation}
\label{eq:B5}
f(\theta = 0)= \frac{k_{\rm out} R^2}{2i} \int\limits_{0}^{1} \left( 1- e ^{i 2(k_{\rm in}-k_{\rm out})R\cos \psi}\right) {\rm d} \cos^2\psi .
\end{equation}

Then, the expression for the total cross section is obtained from Eq.~(\ref{eq:B4}) and setting $w=\cos\psi$, we get
\begin{equation}
\begin{split}
\sigma_{\rm tot}&=\mathop{\mathrm{Im}}\left[ i \,4\pi R^2\int_0^{1}w\left(1 - e ^{ i (k_{\rm in}-k_{\rm out})2Rw}\right) {\rm d} w\right] \\
&=\mathop{\mathrm{Im}}\left[ i \,2\pi R^2- i \, 4\pi R^2\int_0^1 w\, e ^{ i (k_{\rm in}-k_{\rm out})2Rw} {\rm d} w\right].
\end{split}
\label{eq:B6}
\end{equation}

Then, from integration by parts, we obtain
\begin{equation}
\begin{split}
\sigma_{\rm tot}&=\mathop{\mathrm{Im}}\left[ i \,2\pi R^2- i \, 4\pi R^2\left(\left[w\frac{ e ^{ i (k_{\rm in}-k_{\rm out})2Rw}}{ i (k_{\rm in}-k_{\rm out})\,2R}\right]_0^1-\int_0^1\frac{ e ^{ i (k_{\rm in}-k_{\rm out})2Rw}}{ i (k_{\rm in}-k_{\rm out})\,2R} {\rm d} w\right)\right]\\
&=\mathop{\mathrm{Im}}\left[ i \,2\pi R^2- i \, 4\pi R^2\left(\frac{ e ^{ i (k_{\rm in}-k_{\rm out})2R}}{ i (k_{\rm in}-k_{\rm out})\,2R}+\frac{ e ^{ i (k_{\rm in}-k_{\rm out})2R}-1}{(2R)^2\,(k_{\rm in}-k_{\rm out})^2}\right)\right]\\
&=2\pi R^2-4\pi R^2\left[\frac{\sin\left(2R\,(k_{\rm in}-k_{\rm out})\right)}{2R\,(k_{\rm in}-k_{\rm out})}-\frac{1}{2}
\left[ \frac{\sin\left(R\,(k_{\rm in}-k_{\rm out})\right)}{R\,(k_{\rm in}-k_{\rm out})}\right]^2
\right].
\end{split}
\label{eq:B7}
\end{equation}

If we introduce the parameter $\Delta_{\rm R} = 2R\,(k_{\rm in}-k_{\rm out})$, the analytic expression for the cross section becomes
\begin{equation}
\sigma_{\rm tot} = 2\pi R^2\left[1 - 2\,\frac{\sin\Delta_{\rm R}}{\Delta_{\rm R}}+\left(\frac{\sin\Delta_{\rm R}/2}{\Delta_{\rm R}/2}\right)^2\right].
\label{eq:B8}
\end{equation}

In this expression, the total cross section approaches $2\pi R^2$ at high energies. This is the so called ``black disk'' approximation. Nevertheless, from a purely geometrical viewpoint, between two spheres, a collision occurs if the distance between their centres is less than $r_{1}+r_{2}$, where $r_{1}$ and $r_{2}$ are their radii. So the geometrical cross section is equal to $\pi (r_{1}+r_{2})^2$. In our case, the photon may be considered to be a particle with a diameter equal to its reduced wavelength $\bar{\lambda} = \lambda / 2 \pi$~\cite{Narayan}. Consequently, Eq.~(\ref{eq:B8}) becomes
\begin{equation}
\sigma_{\rm tot} = 2\pi \left(R+\frac{\lambda}{4\pi} \right)^2\left[1 - 2\,\frac{\sin\Delta_{\rm R}}{\Delta_{\rm R}}+\left(\frac{\sin\Delta_{\rm R}/2}{\Delta_{\rm R}/2}\right)^2\right].
\label{eq:B9}
\end{equation}

\begin{figure}[t]
\centering
	\includegraphics[width=4.25in]{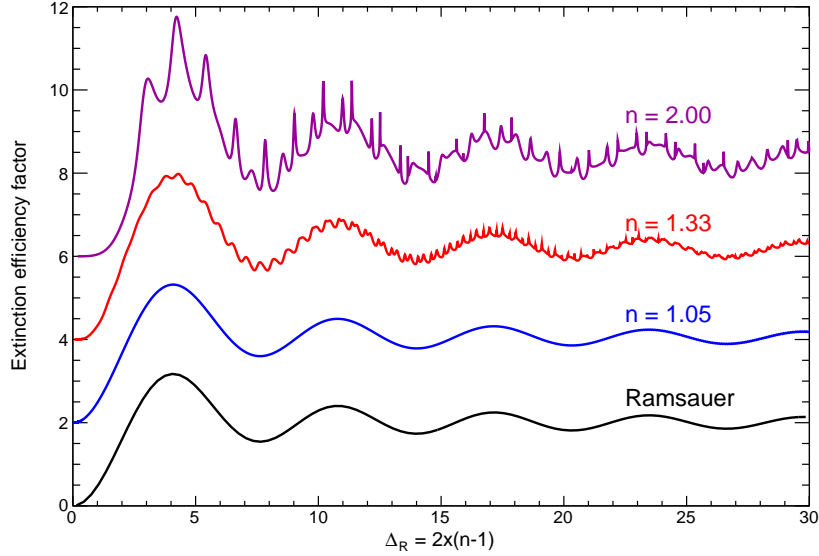}
\caption{{(Color online) The extinction efficiency factor versus the phase shift $\Delta_{\rm R}$, for non-absorbing spherical particles with relative refractive indices $n=1.05,\: 1.33$, and $2.00$}. The Ramsauer solution from Eq.~(\ref{eq:B10}) for $n = 1.05$ is also given. The vertical scale applies only to the lowest curve, the others being successively shifted upward by 2.}
\label{fig:Mie_QextfPhaseShift}
\end{figure}

\begin{figure}[t]
\centering
 \includegraphics[width=4.25in]{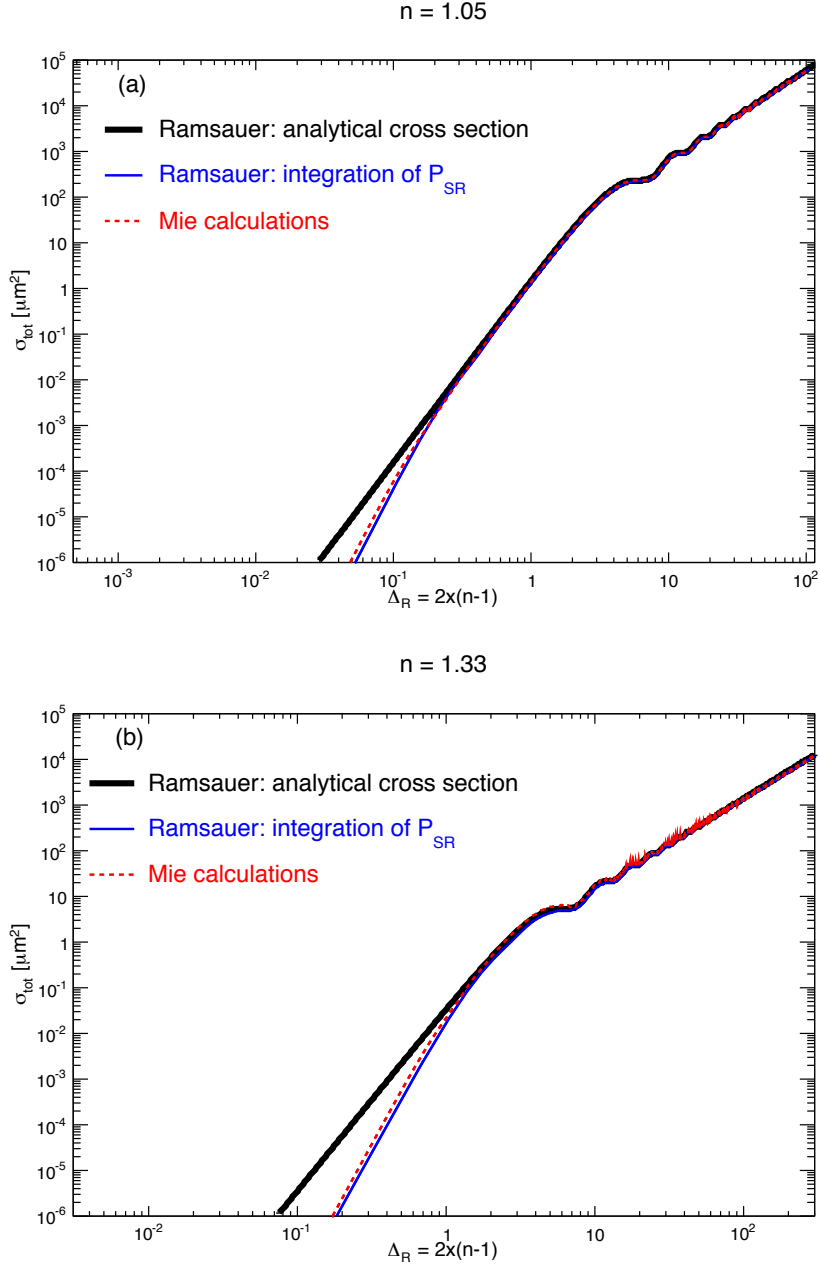}
\caption{{(Color online) Total cross section as a function of the phase shift $\Delta_{\rm R} = 2x(n-1)$ for a refractive index fixed at (a) $n=1.05$ and (b) $n = 1.33$.} The simplified Ramsauer solution from Eq.~(\ref{eq:D1}) is in blue continuous line, the Mie prediction in red dashed line (using~\cite{Laven}) and the formula from Eq.~(\ref{eq:B8}) in black thick line. The incident wavelength is fixed at $0.6~\mu$m.}
\label{fig:ComparisonCS}
\end{figure}

Finally, an expression for $\Delta_{\rm R}$ as a function of the index of refraction $n$ is needed. According to the conservation of phase, it is $k_{\rm in} = n\,k_{\rm out}$. With this relation and the fact that $\lambda = 2 \pi/k_{\rm out}$
\begin{equation}
Q_{\rm ext, R} = \frac{\sigma_{\rm tot}}{\pi R^2} = 2 \left (1+\frac{n-1}{\Delta_{\rm R}} \right)^2 \left[1 - 2\,\frac{\sin\Delta_{\rm R}}{\Delta_{\rm R}}+\left(\frac{\sin\Delta_{\rm R}/2}{\Delta_{\rm R}/2}\right)^2\right],
\label{eq:B10}
\end{equation}
where $\Delta_{\rm R} = 2R\,(k_{\rm in}-k_{\rm out}) = 2R\,(n-1)\,k_{\rm out} = 2x\,(n-1)$. This result is derived using the approximation of scalar light, unlike the Mie solution. Therefore our formula applies only to scattering of unpolarised light. Note that the first factor of Eq.~(\ref{eq:B10}) explains the fact that the amplitude of the large oscillations grows with $n$. The extinction efficiency factors, when plotted against $\Delta_{\rm R}$, show a universal shape (see Fig.~\ref{fig:Mie_QextfPhaseShift}). Our model and the Mie predictions are in good agreement for the three refractive indices used.

\subsection{Ramsauer total cross section via the differential cross section}
The simplified Ramsauer formula is very convenient to use, in computing the phase function, as we have just seen, and also in calculating the total cross section. In this section, we compare the Mie prediction with the one obtained by integration of Eq.~(\ref{eq:C11}). From the differential cross section, the total cross section is given by
\begin{equation}
\sigma_{\rm tot} = \int\limits_0^{2\pi}{\rm d} \Phi \int\limits_0^\pi  {\rm d} \theta \sin \theta \; \frac{{\rm d} \sigma}{{\rm d} \theta} .
\label{eq:D1}
\end{equation}

Fig.~\ref{fig:ComparisonCS} compares, for two values of the index of refraction, the integrated Ramsauer phase function and the Mie theory result. The analytical solution from Eq.~(\ref{eq:B8}) is also plotted. Each curve is characterised by a succession of maxima and of minima. While there is generally a good agreement between the three curves, only the integration of the simplified Ramsauer solution $P_{\rm SR}$ reproduces the Mie curve at small phase shift $\Delta_{\rm R}$, the so-called Rayleigh regime. The agreement improves when the refractive index decreases. In fact, at low phase shift, i.e.\ at small size parameter, the behaviour of Eq.~(\ref{eq:B8}) is not proportional to $\lambda^{-4}$ as expected by the Rayleigh theory, but only to $\lambda^{-2}$. Of the two methods, only the integration of the simplified Ramsauer solution is able to reproduce correctly the physics of Rayleigh regime.

\section{Application to the Henyey-Greenstein scattering phase function}
\label{sec:HenyeyGreenstein}
Monte Carlo simulations are usually used to describe light scattering in a medium. It is a very global problematic encountered in astrophysics~\cite{HG_astro} as well as in meteorology~\cite{HG_meteo} or biology~\cite{HG_bio}. One of the critical points is to simulate the multiple scattering of photons in the medium. Also, the size distribution of particles or their shape are often unknown: the Mie phase function has to be evaluated for each particle. One of the most popular scattering phase functions is the Henyey-Greenstein (HG) function~\cite{HenyeyGreenstein}. This function was first introduced by Henyey and Greenstein in 1941 to describe scattering processes in galaxies. This is a parameterisation usually used to reproduce scattering on objects large with respect to the incident wavelength. In the case where the backscattering can not be neglected, the HG function becomes a ``Double HG'' and is given by
\begin{equation}
P_{\rm HG}(\theta|g_{\rm HG},f) =  \frac{1-{g_{\rm HG}}^2}{4\pi}\left[\frac{1}{\left(1+{g_{\rm HG}}^2-2g_{\rm HG}\cos\theta\right)^{3/2}} + f\frac{3\cos^2\theta-1}{2\left(1+{g_{\rm HG}}^2\right)^{3/2}} \right],
\label{eqHG_1}
\end{equation}
where $\theta$ is the scattering angle, $f$ represents the strength of the second component relative to the backward scattering peak ($f=0$ meaning no additional component) and $g_{\rm HG}$ is the so-called asymmetry parameter. It is defined as
\begin{equation}
g_{\rm HG}=\langle \cos\theta \rangle = \int\limits_0^\pi \cos\theta\,P_{\rm HG}(\theta|g_{\rm HG},f)\,2\pi\sin\theta \,{\rm d}\theta .
\label{eqHG_2}
\end{equation}

\begin{figure}[t]
\centering
	\includegraphics[width=4.25in]{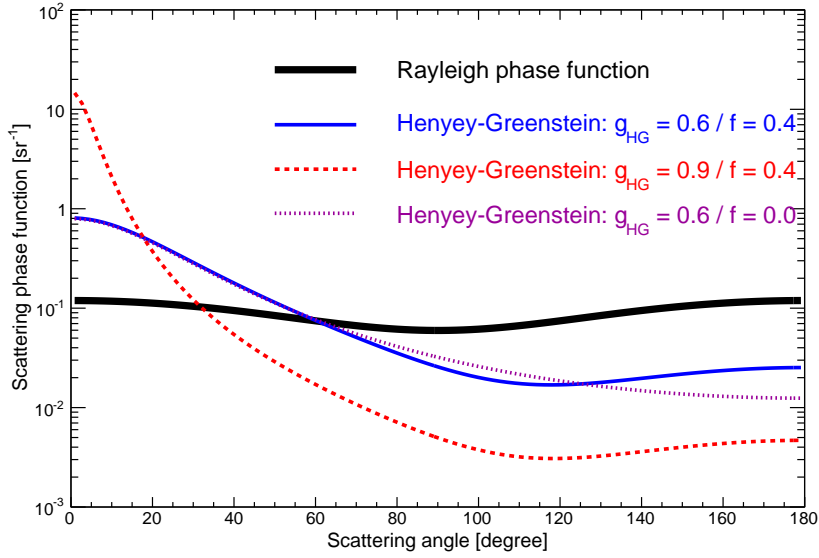}
\caption {\label {fig:PhaseFunctionHG}{(Color online) Henyey-Greenstein functions representing the scattering phase function for different asymmetry parameters $g_{\rm HG}$ and backward factors $f$.} The Rayleigh phase function, proportional to $(1+\cos^2 \theta)$, is also plotted.}
\end{figure}

\begin{figure}[t]
\centering
	\includegraphics[width=4.25in]{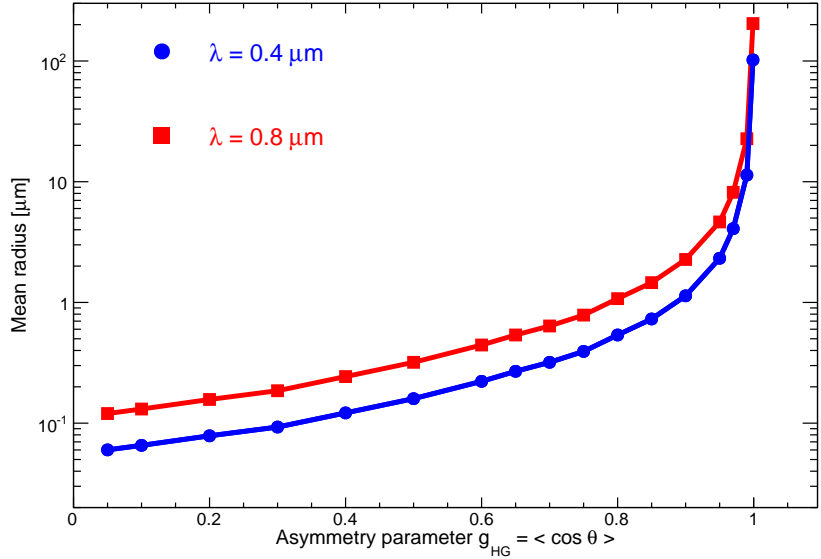}
\caption{\label {fig:EquivalenceHG_size} (Color online) Relation between the asymmetry parameter of the Henyey-Greenstein function and the mean radius of the particle size distribution. The equivalence is plotted for two different incident wavelengths: $0.4~\mu$m (blue circles) and $0.8~\mu$m (red squares).}
\end{figure}

Henyey-Greenstein functions for different values of $g_{\rm HG}$ and $f$ are plotted in Fig.~\ref{fig:PhaseFunctionHG}. The asymmetry factor affects the strength of the forward peak and its width. Contrary to the Rayleigh phase function, Henyey-Greenstein phase functions depict a strong forward peak directly linked to the asymmetry parameter $g_{\rm HG}$. Using the Ramsauer approach developed in Sect.~\ref{sec:ramsauer} and the resulting relation $\Delta \theta_{\rm FWHM} = 2/k_{\rm out}R$, the mean radius of the particle size distribution can be estimated from the width of the forward peak, for a fixed incident wavelength. Fig.~\ref{fig:EquivalenceHG_size} shows the relation between $g_{\rm HG}$ values and mean particle radius for two different incoming wavelengths $\lambda$. To take into account the overestimation of the mean radius explained in Sect.~\ref{sec:Ramsauer_phasefunction}, the values plotted have been reduced of $30\%$. In some cases, only the asymmetry parameter is measured by an experimental setup: these curves can be useful to estimate directly a mean particle size for the medium studied. If the first goal is to estimate the particle size distribution, a more detailed analysis seems to be needed. However, this method gives a first estimation of the particle size and an intuitive understanding of the asymmetry parameter $g_{\rm HG}$ for the Henyey-Greenstein parameterisation.

It is typical of what is needed in large ground-based astrophysics projects as cosmic ray shower detectors~\cite{PAO} or imaging atmospheric Cherenkov telescopes~\cite{CTA}. In these experiments, the atmosphere is used as a giant calorimeter and the associated systematic uncertainties have to be as small as possible~\cite{AugerATMO_LongPaper}. Usually, atmospheric effects are divided into two categories: the molecular component~\cite{EPJP_BiancaMartin}, monitored using ground-based weather stations and meteorological radio soundings, and the aerosol component~\cite{EPJP_Aerosol}. Atmospheric aerosols are small particles as dust or droplets in suspension having a typical size varying from a few nanometres to a few micrometres. Most of the atmospheric aerosols are present only in the first few kilometres above the ground. Unlike the molecular component, the aerosol population is highly variable in time and location, depending on the wind and weather conditions. Two main physical quantities have to be estimated to correct the effect of the aerosols on the number of photons detected by the telescopes: the aerosol attenuation length and the aerosol scattering phase function. Some of these astrophysics projects have already published an estimation of the asymmetry parameter~\cite{Benzvi,MyICRC}, giving a first estimation of the mean aerosol size present in the low part of the atmosphere.

\section{Conclusion}
We have shown that within the experimental errors the Ramsauer method predicts an extinction efficiency factor and a scattering phase function compatible with those given by the Mie solution. The large oscillatory behaviour for the extinction efficiency factor is understood as the consequence of the interference between the fraction of light going through and the fraction passing close to the sphere.

Two Ramsauer solutions have been given for the scattering phase function. The form of the first one, $P_{\rm R}(\theta)$, without approximations, itself does not lead to obvious physical interpretation. The second, the simplified Ramsauer solution $P_{\rm SR}(\theta)$, is an analytical formula from which characteristic behaviour can be predicted. However it cannot reproduce the Rayleigh regime for small size parameters.

The Henyey-Greenstein function is a parameterisation intensively used in Monte Carlo simulations describing light scattering. Its key parameter is the asymmetry parameter $g_{\rm HG}$. Using the Ramsauer approach, we proposed a basic relation between $g_{\rm HG}$ and the mean radius of particles scattering light. This approach is already employed and seems to be useful in large astrophysics projects.

\section*{Acknowledgments}
The authors would like to thank Paul Eschstruth, Darko Veberi\v{c} and the three referees for fruitful discussions and/or for their comments on the manuscript. 

\newpage


\end{document}